\documentstyle[epsfig]{mn}
\ifCUPmtlplainloaded \else

\fi
\title[The X-ray Transient XTE J2012+381] {The X-ray Transient XTE J2012+381}
\author[R. I. Hynes et al.]
       {R.I.~Hynes$^1$,
       P. Roche$^1$\thanks{\protect\raggedright
       Present address: Space Research Centre, Department of Physics 
       and Astronomy, University of Leicester, LE1 7RH}, 
       P. A. Charles$^2$, M. J. Coe$^3$\\
$^1$Astronomy Centre, University of Sussex, Falmer, Brighton BN1 9QJ\\
$^2$Astrophysics, Nuclear and Astrophysics Laboratory,
    Keble Road, Oxford, OX1 3RH\\
$^3$Department of Physics and Astronomy, University of Southampton, 
    Southampton, SO17 1BJ}
\begin{document}
%
%
\newcommand{\novacyg}{V404~Cyg}
\newcommand{\novasco}{GRO\,J1655--40}
\newcommand{\novaper}{GRO\,J0422+32}
\newcommand{\novamus}{X-ray Nova Muscae 1991}
\newcommand{\novamon}{A\,0620--00}
\newcommand{\novavul}{GS\,2000+25}
\newcommand{\xtesource}{XTE\,J2012+381}
%
%
\newcommand{\HST} {\textit{HST}}
\newcommand{\XTE} {\textit{RXTE}}
\newcommand{\RXTE}{\textit{RXTE}}
\newcommand{\GRO} {\textit{GRO}}
\newcommand{\ASCA}{\textit{ASCA}}
%
%
\newcommand{\HI}   {H\,\textsc{i}}
\newcommand{\HII}  {H\,\textsc{ii}}
\newcommand{\HeI}  {He\,\textsc{i}}
\newcommand{\HeII} {He\,\textsc{ii}}
\newcommand{\HeIII}{He\,\textsc{iii}}
\newcommand{\MgII} {Mg\,\textsc{ii}}
%
%
\newcommand{\EBV}{E_{{\rm B} - {\rm V}}}
\newcommand{\EVR}{E_{{\rm V} - {\rm R}}}
\newcommand{\Rv} {R_{\rm V}}
\newcommand{\Av} {A_{\rm V}}
%
%
\newcommand{\lam}   {$\lambda$}
\newcommand{\lamlam}{$\lambda\lambda$}
\maketitle
%
%
\newcommand{\comm}[1]{\textit{[#1]}}
%
%
\begin{abstract}
We present optical and infrared observations of the soft X-ray
transient (SXT) \xtesource\ and identify the optical counterpart with
a faint red star heavily blended with a brighter foreground star.  The
fainter star is coincident with the radio counterpart and appears to
show weak H$\alpha$ emission and to have faded between observations.
The \RXTE /ASM lightcurve of \xtesource\ is unusual for an SXT in that
after an extended linear decay, it settled into a plateau state for
about 40 days before undergoing a weak mini-outburst.  We discuss the
nature of the object and suggest similarities to long orbital period
SXTs.
\end{abstract}
%
%
%
%
\section{Introduction}
\begin{figure}
\begin{center}
\epsfig{width=2in,height=3in,angle=90,file=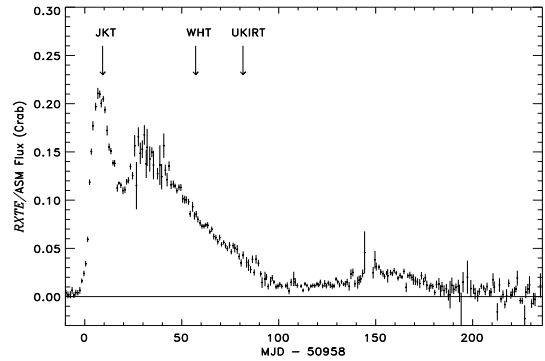}
\caption{\RXTE /ASM one-day average light curve based on quick-look
results provided by the \RXTE /ASM team.  The zero point in time is
chosen to coincide approximately with the first reported \RXTE\
detection on MJD 50957.6 (Remillard, Levine \& Wood 1998).  Times of
JKT, WHT and UKIRT observations are marked.  Note that a mini-outburst
occurs around day 145.}
\label{ASMFig}
\end{center}
\end{figure}

The transient X-ray source \xtesource\ was discovered by the \RXTE\
All Sky Monitor (ASM) on 1998 May 24 \cite{RLW98}.  Its ASM light
curve is shown in Fig.\ \ref{ASMFig}.  \ASCA\ observations (White et
al.\ 1998) revealed the ultrasoft spectrum plus hard power-law tail
signature of a black hole candidate.  A radio counterpart was
suggested by Hjellming \& Rupen \shortcite{HR98a} and found to be
close to an 18th magnitude star (Castro-Tirado \& Gorosabel 1998,
Garcia et al.\ 1998), the USNO A1.0 star 1275.13846761 \cite{M96}.
Spectroscopy of this star showed a nearly featureless spectrum with
Balmer and Na~D absorption, and no apparent emission lines (Garcia et
al.\ 1998).  Our images obtained with the Jacobus Kapteyn Telescope
(JKT) on La Palma, however, showed the presence of a faint red
companion star 1.1\,arcsec away, closely coincident with the radio
source (Hynes and Roche 1998, Hjellming and Rupen 1998b).  Infrared
images also showed this second star \cite{C98}.

In this paper, we report in detail on the JKT photometry of the faint
red star, which we suggest to be the true optical counterpart of
\xtesource.  A spectrum obtained with the William Herschel Telescope
(WHT) reveals weak H$\alpha$ emission in the fainter star supporting
its identification with \xtesource.  The red star also appears
somewhat fainter at the epoch of the WHT observations.  Infrared
imaging obtained with the United Kingdom Infrared Telescope (UKIRT)
late in the X-ray decline clearly shows both stars and also suggests
fading relative to the earlier observations of Callanan et al.\
(1998).
%
%
\section{JKT Photometry}
\begin{figure}
\fbox{\epsfig{width=3.2in,file=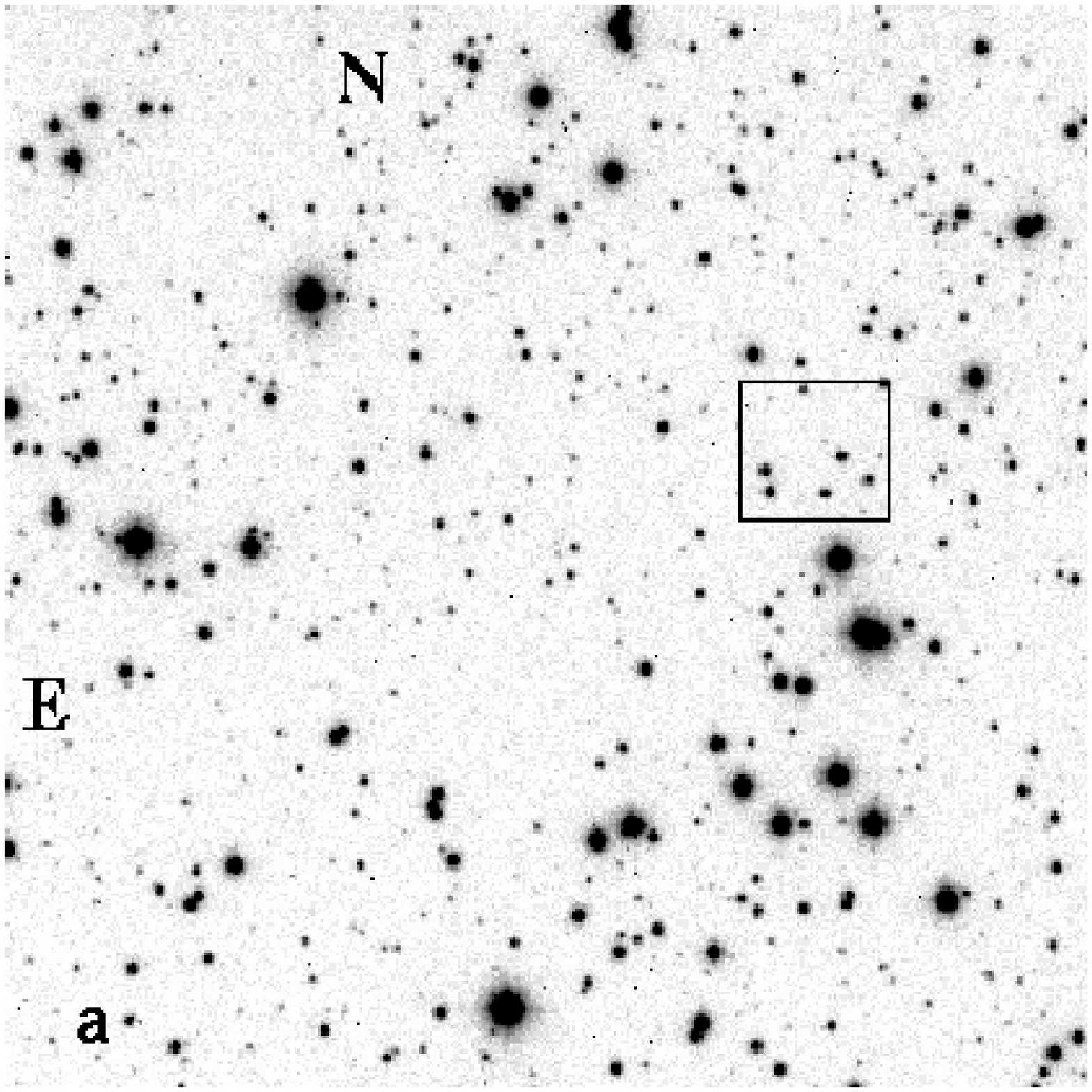}}
\newline
\vspace*{-6pt}
\newline
\fbox{\epsfig{width=24.8mm,file=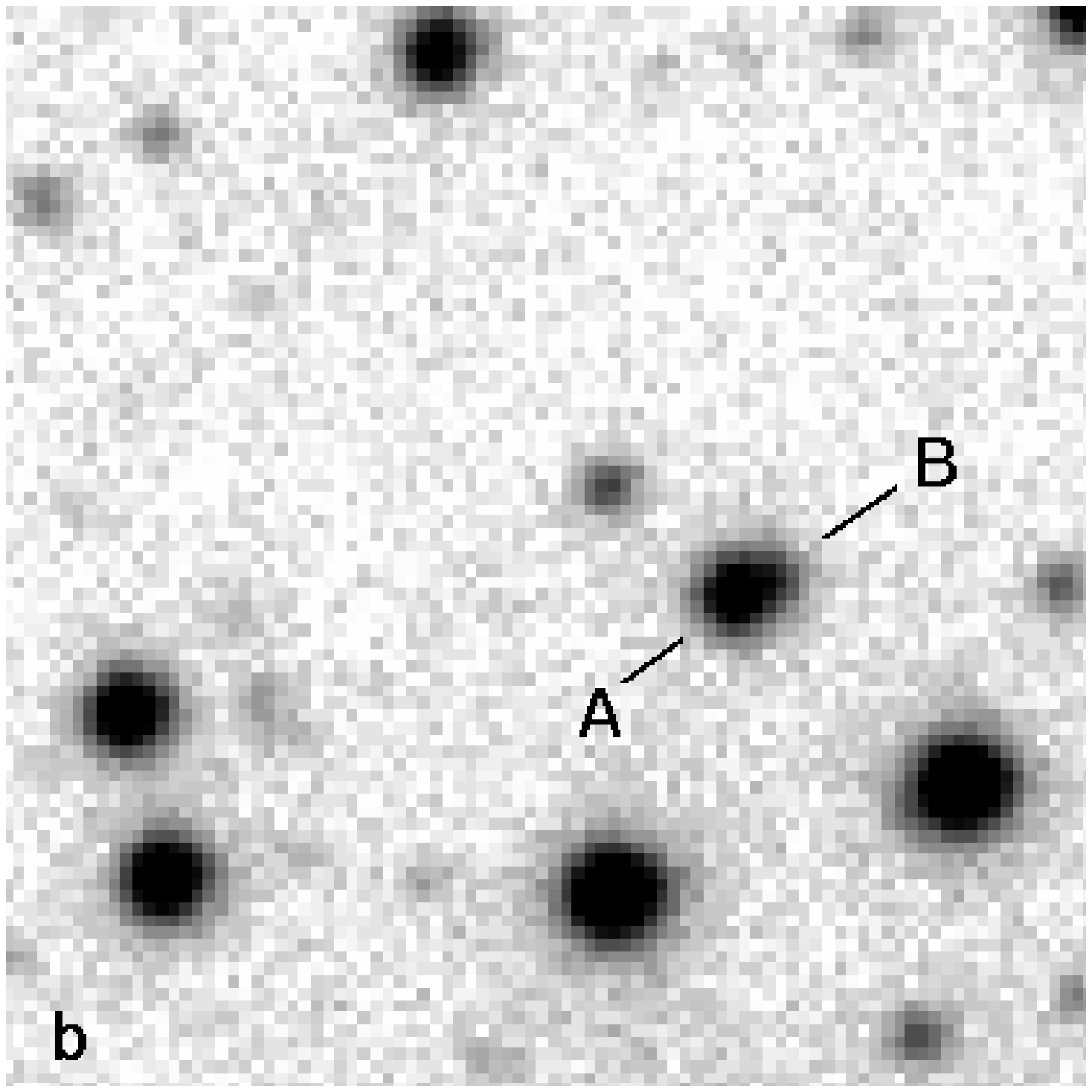}}
\fbox{\epsfig{width=24.8mm,file=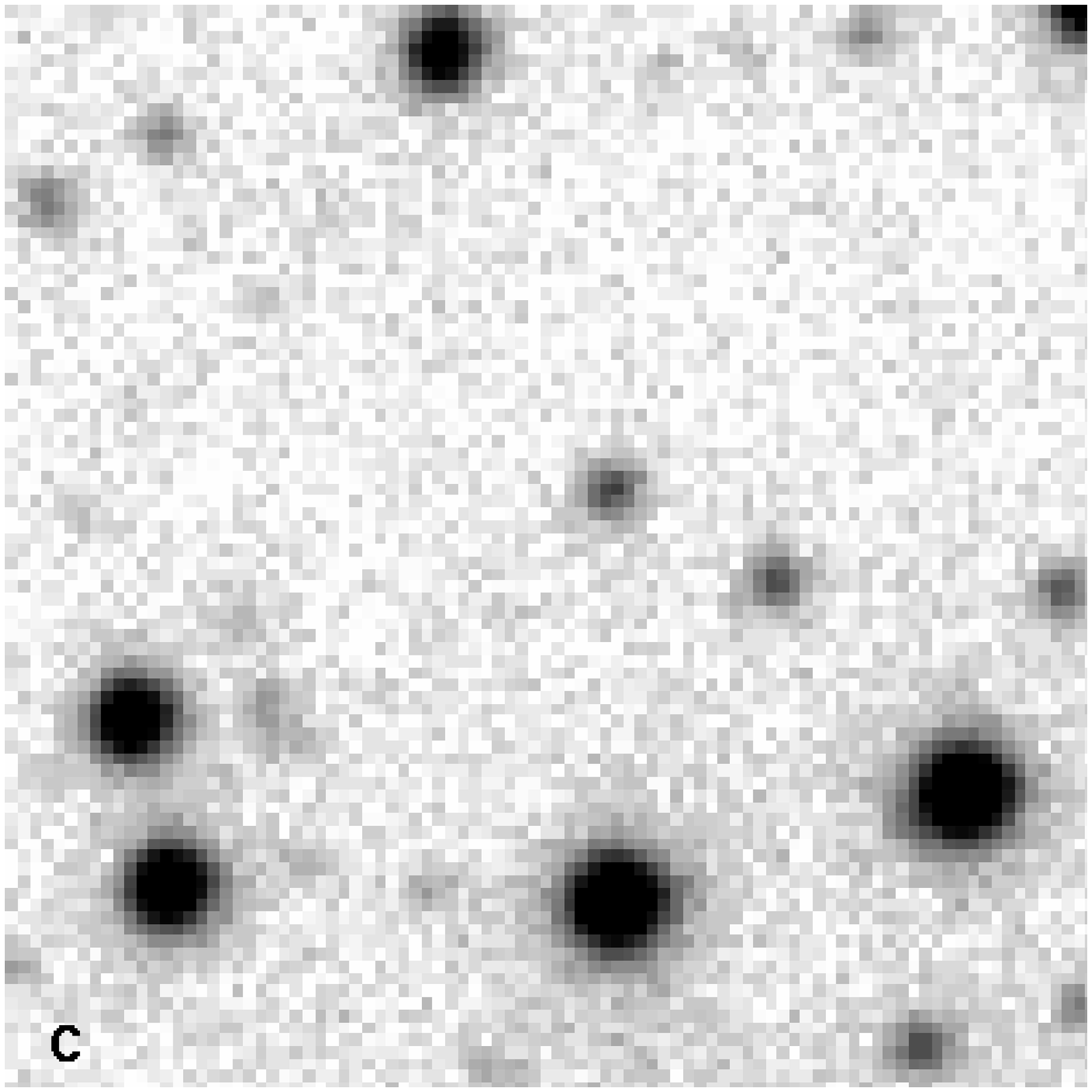}}
\fbox{\epsfig{width=24.8mm,file=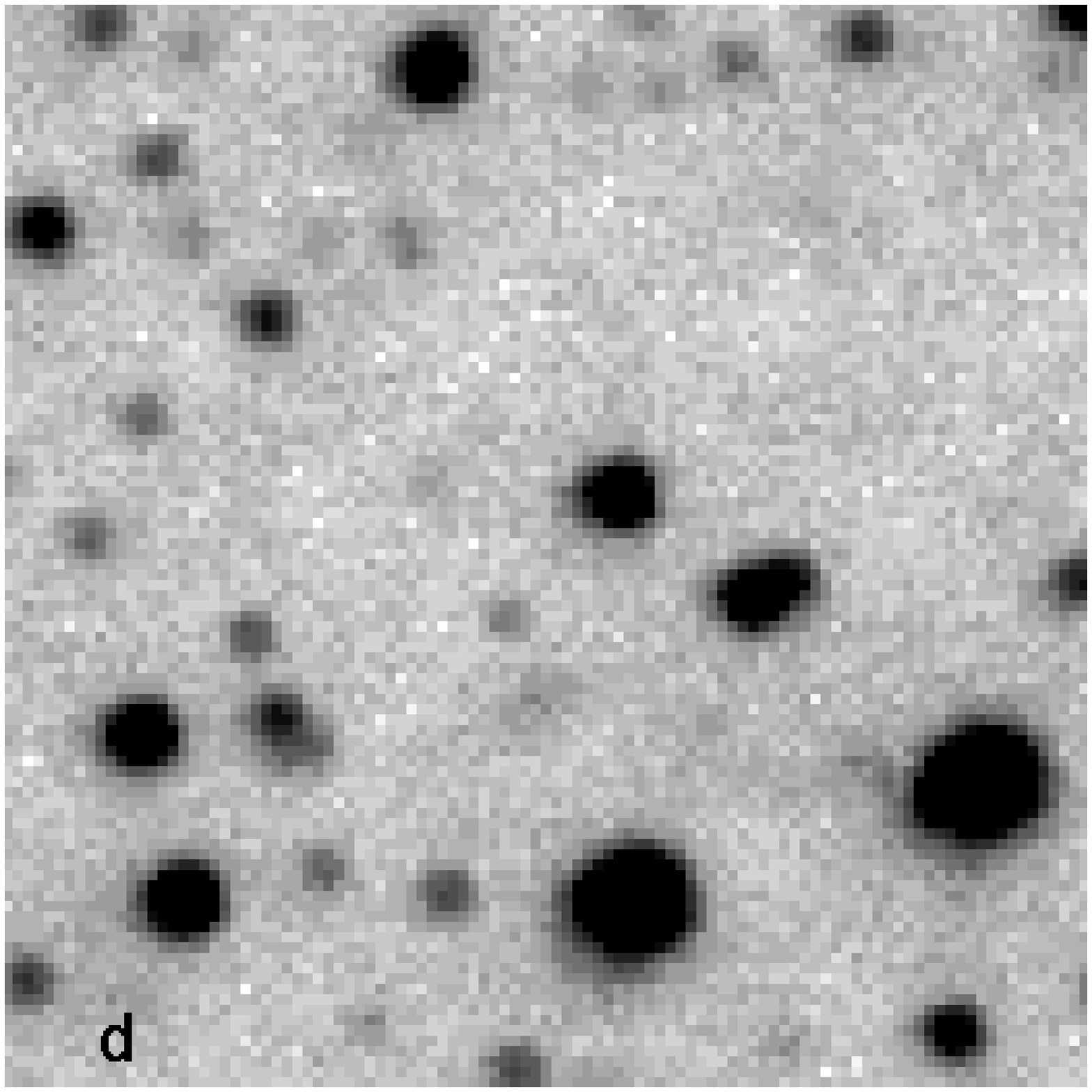}}
\caption{a) V band finding chart for \xtesource (4-arcmin square).
The boxed area is blown up as b--d.  b,c) Expanded I band JKT image
before and after subtraction of the USNO star (A) from 1998 June 3.
The fainter star (B) is just detectable before subtraction and clearly
remains after.  d) K band image from 1998 August 14.}
\label{PSFFig}
\end{figure}

Multicolour photometry of the field of \xtesource\ was taken on 1998
June 3 through the JKT service programme.  The JKT CCD camera was used
with the TEK4 CCD and a standard UBVRI filter set.  De-biasing and
flat-fielding were performed with standard \textsc{iraf} tasks.  A V
band image of the field is shown in Fig.\ \ref{PSFFig}a.  As the field
is crowded, the \texttt{daophot} task was used to deblend point spread
functions.  I band images before and after subtraction of the PSF of
USNO A1.0 star 1275.13846761 (Monet et al.\ 1996; hereafter the USNO
star) are shown in Fig.\ \ref{PSFFig}b and \ref{PSFFig}c.  A residual
stellar image is clearly present after subtraction.  Its offset
relative to the USNO star was measured using V, R and I images; the
results are consistent to 0.1\,arcsec.  Hence we determine the position
of the star (Table \ref{PosTable}).  The position of the radio source
(Hjellming \& Rupen 1998b) is consistent with the fainter star to
within uncertainties, but is difficult to reconcile with the USNO
star.  The fainter star is therefore a strong candidate for the
optical counterpart of the radio source (and hence the X-ray source).

Absolute flux calibration was obtained from a colour-dependent fit to
five stars from Landolt standard field 110 \cite{L92}.  The fainter
star could only be distinguished in V, R and I band images and its
magnitudes (Table \ref{MagTable}) indicate that it is very red, with
${\rm V} - {\rm R} = 1.4\pm 0.2$.  Our only quantitative estimate of
the interstellar reddening comes from White et al.\ \shortcite{W98}
who estimate a column density $N_{\rm H} = (1.29 \pm 0.03) \times
10^{22}$\,cm$^{-2}$.  Gas-to-dust scalings are at best approximate.
Using the relations of Ryter, Cesarsky \& Audouze (1975), Bohlin,
Savage \& Drake (1978) and Predehl \& Schmitt (1995) we obtain $\EBV =
1.9$, 2.2 and 2.4 respectively, so we adopt this range as a reasonable
estimate of reddening.  Assuming the extinction curve of Cardelli,
Clayton \& Mathis (1989) this implies $1.4 < \EVR < 1.8$ and hence an
intrinsic colour of $-0.6 < {\rm V} - {\rm R} < 0.2$.  This is
consistent with other SXTs; the optical emission is likely dominated
by a hot accretion disc.  This colour would also be consistent with an
early type star, but the lack of H$\alpha$ absorption (Sect.\
\ref{WHTSect}) would not, unless it were almost exactly filled in by
emission.

%
\begin{table}
\caption{Positions of the USNO A1.0 star 1275.13846761 (Monet et al.\
1996), the faint red star we identify with \xtesource, and the radio
source (Hjellming \& Rupen 1998b).}
\label{PosTable}
\begin{tabular}{lccc}
             & $\alpha$ & $\delta$ & Error\\
USNO star & 
$20^{\rm h}12^{\rm m}37\fs80$ &
$38\degr11\arcmin00\farcs6$ &
$0\farcs25$ \\
Fainter star & 
$20^{\rm h}12^{\rm m}37\fs71$ &
$38\degr11\arcmin01\farcs1$ &
$0\farcs35$ \\
Radio source & 
$20^{\rm h}12^{\rm m}37\fs67$ &
$38\degr11\arcmin01\farcs2$ &
$0\farcs4$ \\
\end{tabular}
\end{table}
\begin{table}
\caption{VRI magnitudes of the two stars closest to the radio position
of \xtesource\ and the star used as a comparison for the WHT spectrum
(Sect.\ \ref{WHTSect}.)  JKT data from 1998 June 3.}
\label{MagTable}
\begin{tabular}{lccc}
              & V & R & I \\
USNO star     & $17.91\pm0.05$ & $17.15\pm0.05$ & $16.51\pm0.05$ \\
Red companion & $21.33\pm0.10$ & $19.90\pm0.15$ & $18.64\pm0.10$ \\
Comparison    & $17.69\pm0.05$ & $16.84\pm0.05$ & $16.12\pm0.05$ \\
\end{tabular}
\end{table}
%
%
\section{WHT Spectroscopy}
\label{WHTSect}
\xtesource\ was observed with the WHT on 1998 July 20, when the object
had faded to approximately half its peak X-ray brightness.  The ISIS
dual-beam spectrograph was used in single red arm mode to maximise
throughput with the low-resolution R158R grating
(2.9\,\AA\,pixel$^{-1}$) and the TEK2 CCD.  A 1\,arcsec slit was used
giving an instrumental resolution of 5.5\,\AA.  Based on positions
derived from the JKT photometry, the slit was aligned to pass through
the line of centres of the USNO star and an isolated comparison star
used to define the spatial profile as a function of wavelength.  This
comparison star was chosen to lie along the line of centres of the two
blended stars.  The three stars are co-linear to within the accuracy
of our astrometry, i.e.\ the derived V, R and I positions of the red
star scatter evenly to either side of the line of centres of the USNO
star and comparison star.  We estimate that the offset of the red star
from the line of centres of the other two stars is therefore less than
0.1\,arcsec, the scatter in the astrometry.  The positioning of the
slit was judged by eye using reflections off the slit jaws.  There is
therefore some uncertainty in centring the stars within the slit, but
as the stars are co-linear to within a fraction of the slit width,
this should not affect their {\em relative} brightness significantly.
Standard {\tt iraf} procedures were used to de-bias and flat-field the
spectrum.  Sky subtraction was performed by fitting fourth order
polynomials to the sky background with the stellar profiles masked
out.  The subtracted image showed no significant residuals in the sky
regions.  Wavelength calibration was achieved using a fit to a
copper-neon arc spectrum obtained immediately before the object
spectrum and checked against the night sky emission lines.

\begin{figure}
\begin{center}
\epsfig{height=3in,angle=90,file=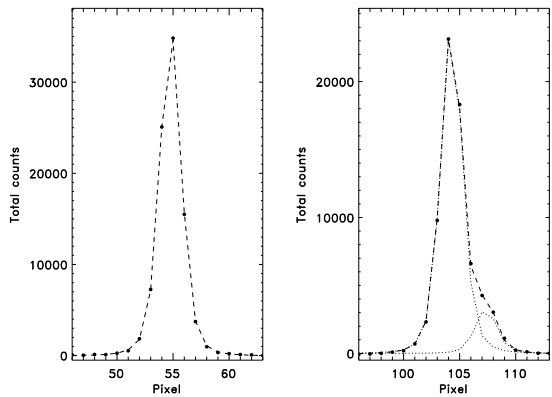}
\caption{Fit to spatial profiles in the red part of the WHT spectrum.
Data is shown by points, the model profile is dashed and the dotted
lines indicate the De-blended components.  Both data and fits have been
summed over 20 pixels in the dispersion direction to reduce noise and
illustrate the quality of fit achieved.  On the left is the fit to the
spatial profile of the template star.  On the right is the two-profile
fit to the stars of interest.}
\label{SpatialFig}
\end{center}
\end{figure}

Although the seeing was good (spatial FWHM 0.9\,arcsec) the stars are
still heavily blended; a sample binned spatial cut is shown in Fig.\
\ref{SpatialFig}.  The third star on the slit was therefore used to
define the spatial profile as a function of wavelength, by means of a
Voigt profile fit.  The profile parameters are smoothed in wavelength
using a fourth-order polynomial fit and then fixed.  The centre
position of the profile varies by less than one pixel over the whole
spectrum.  The Gaussian core width and the Voigt damping parameter
both vary by about 5\,per cent.  No smaller scale patterns in the
residuals to this fit are apparent.  The Voigt model was found to give
a very good fit over most of the profile.  It does somewhat
overestimate the extreme wings of the profile, so we approximately
correct this by defining a wavelength independent correction term as a
function only of distance from profile centre.  With this correction,
no significant deviations between model and data can be discerned (see
Fig.\ \ref{SpatialFig}).  The positions of the other two stars
relative to the isolated comparison are taken as fixed, leaving only
the amplitudes of two blended profiles to be fitted.  This was done by
$\chi^2$ minimisation and is effectively a generalisation of the
optimal extraction method \cite{H85} to the multiple profile case.
This method for separating the spectra of blended stars will be
described in more detail in a subsequent paper in preparation.

A Telluric correction spectrum was formed by normalising the spectrum
of the spectrophotometric standard BD+40 4032 \cite{S77} observed
later in the night.  This has spectral type B2\,III so H$\alpha$ and
\HeI\ absorption lines were masked out.  The correction spectrum was
shifted slightly to compensate for spectrograph flexure between this
exposure and that of our target.  It was also rescaled to give a
least-squares fit to the normalised sum of the three spectra on the
slit.  A short exposure of our target was made immediately before
observing BD+40~4032 so we also use this to obtain absolute flux
calibration.  Both these calibration observations used a wide slit
(4\,arcsec) to ensure photometric accuracy.

\begin{figure}
\begin{center}
\epsfig{height=3in,angle=90,file=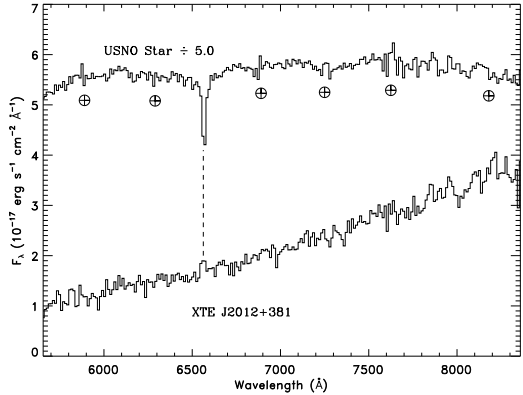}
\caption{WHT spectra of the USNO star (early F type, Garcia et al.\
1998) and the faint red companion believed to be the optical
counterpart of \xtesource.  Both have been binned $\times4$ in
wavelength for clarity.  Atmospheric absorption features have been
corrected for in both spectra.  The approximate locations of the
strongest features are marked $\oplus$ to indicate where residuals
from this correction may distort the spectrum.  The only prominent
line in either spectrum is H$\alpha$: absorption in the brighter star
and weak emission in the companion.}
\label{SpecFig}
\end{center}
\end{figure}

The final calibrated spectra of both stars are shown in Fig.\
\ref{SpecFig}, binned $\times4$ in wavelength.  As can be seen, the
Telluric corrections are good, though slight distortion of the
spectrum remains, e.g.\ near 7250\,\AA.  The only prominent
non-atmospheric feature in either spectrum is H$\alpha$, in absorption
in the USNO star (believed to be an early F-type star, Garcia et al.\
1998), with an equivalent width of $(6.5\pm0.3)$\,\AA, and apparent
weak emission in its red companion, equivalent width
$(-6.6\pm1.5)$\,\AA.  The errors in equivalent width are statistical
errors derived from the residuals of a low-order fit to the
surrounding continuum.  Assuming these errors are correct, our
detection of H$\alpha$ is significant at the $3\sigma$ level.

The emission feature appears roughly square, with full width $40\,{\rm
\AA}\sim1800$\,km\,s$^{-1}$.  It is broader than the absorption in the
USNO star (FWHM$\sim 10$\,\AA) and so does not look like a simple
reflection of the absorption line.  It is also much wider than the
night sky H$\alpha$ emission.  We have repeated the extraction process
without the correction to the profile wings.  While this noticeably
degrades the profile fit, it does not affect the H$\alpha$ emission
feature.  As a final test, we have used the derived spatial profile
(including profile correction) to synthesise a new fake image in which
the spectrum of the fainter star was left unchanged, but the H$\alpha$
line profile of the USNO star was used as a template to add artificial
absorption lines (of the same strength and width) at 6100, 6425 and
6725\,\AA.  We then repeated the extraction process on this fake image
without allowing the profile correction to the line wings, i.e.\ using
a model for the spatial profile which is known to be inadequate.  This
was done as a check that a combination of misfitting the spatial
profile and a strong absorption line does not produce spurious
emission.  No emission features are seen in the spectrum of the
fainter star at the position of the fake absorption lines.  After
subtraction of a low-order fit to the continuum, we measure the total
residual counts in 15-pixel (44\,\AA) bins centred on each of the line
positions.  From the noise in the surrounding continuum, we estimate
that the error on the total counts from such a bin is 130 counts.  At
6100, 6425 and 6725\,\AA\ respectively we measure 150, 76 and $-180$
counts, a distribution (mean 14, rms 140) consistent with zero to
within the error estimate.  For the bin centred on H$\alpha$, however,
we measure 550 counts, significantly larger than the estimated error.

We therefore conclude that the H$\alpha$ emission is unlikely to be an
artifact of the de-blending process but represents real emission from
the optical counterpart of the X-ray source, detected with $3\sigma$
confidence.
%
%
\section{UKIRT Photometry}
\label{UKIRTSect}
UKIRT observations were obtained on 1998 August 14. The weather was
clear, winds were light and the humidity was low. The images were
obtained using the IRCAM3 detector. The $256\times256$ detector array
has a scale of 0.286\,arcsec pixel$^{-1}$. The data on our target were
collected between 12.51 and 13.57 UT, during which the airmass varied
from 1.49 to 2.02. Initial data reduction was performed using
\textsc{ircamdr} software.  As for the JKT observations, point spread
function fitting was done using \texttt{daophot} in \textsc{iraf}.
Flux calibration was performed with respect to the UKIRT standard
FS4. The primary data on the standard were taken between the J and K
band observations of the source.  These were compared with other
standard observations taken at the beginning of the sequence as a
check of consistency; the scatter in standard measurements is
$\sim0.03$\,mag.  Our magnitudes for the two stars are summarised in
Table \ref{IRTable} together with the magnitude equivalent to the sum
of their fluxes.  Error estimates are derived from a combination of
the consistency of standard measurements (as an indicator of the
reliability of the calibration) and statistical errors determined by
\texttt{daophot}.
\begin{table}
\caption{J and K magnitudes of the two stars closest to the radio
position of \xtesource\ (Sect.\ \ref{UKIRTSect}).  UKIRT data from
1998 August 14.}
\label{IRTable}
\begin{tabular}{lcc}
              & J              & K              \\
USNO star     & $15.50\pm0.03$ & $15.00\pm0.03$ \\
Red companion & $17.26\pm0.04$ & $16.11\pm0.04$ \\
Combined      & $15.30\pm0.03$ & $14.67\pm0.03$ \\
\end{tabular}
\end{table}
%
%
\section{Is the Optical/IR Counterpart Fading?}
In order to assess whether the optical counterpart had faded between the
JKT and WHT observations, we formed a model bandpass (based on filter
transmission, CCD response and the extinction curve for the same
airmass as the JKT observations).  This was applied to our spectra of
the three stars on the slit to produce synthetic R band magnitudes for
comparison with the photometry.  We estimate ${\rm R}=17.23$ for the
USNO star, ${\rm R}=20.23$ for the faint optical counterpart to the
X-ray source and ${\rm R}=16.89$ for the isolated star.  Our spectra
do not quite cover the full R bandpass, but the loss is a very small
amount at both short and long wavelengths.  We estimate that the
effect of this, together with errors in the model bandpass introduces
an uncertainty of no more than 0.05 magnitude when combined with the
colour differences between the stars.  The JKT and WHT observations of
the USNO star and the slit comparison star are then approximately
consistent, with perhaps a small systematic calibration error of less
than 0.10 magnitudes.  The much larger difference in the magnitudes
for the X-ray source (0.33 magnitudes fainter at the second
observation) then suggests that it has indeed faded optically between
days 9 and 53.

The infrared counterpart also appears fainter than earlier in
outburst.  Callanan et al.\ (1998) estimated ${\rm J}=15.0\pm0.1$,
${\rm K}=14.3\pm0.1$ near the peak of outburst for the two blended
stars.  The comparable combined magnitudes from our UKIRT observation
are given in Table \ref{IRTable}.  Our observations are significantly
different from those of Callanan et al.  To produce this difference by
fading of the fainter star would require a decline of 1.2 magnitudes
in J and 1.0 magnitudes in K between days 9 and 82.
%
%
\section{Discussion}
The X-ray light curve of \xtesource\ (Fig.\ \ref{ASMFig}) shows many
similarities to those of other soft X-ray transients \cite{CSL97}, but
there are some important differences.  The secondary maximum peaks
around day 29, only 22 days after the first peak.  This is earlier in
the outburst than in most systems (45--75 days, Chen, Livio \& Gehrels
1993), but is not unprecedented (Shahbaz, Charles \& King 1998).  The
decline from secondary maximum is clearly linear.  In the paradigm of
King \& Ritter \shortcite{KR98}, this would imply that the disc is too
large to be held in a high state by X-ray irradiation.  A large disc
would then imply a long orbital period.  Such an interpretation is
supported by observations of other SXTs (Shahbaz et al. 1998), in
which extended linear decays are only seen in long-period systems.

A long orbital period in turn implies that XTE J2012+381 probably
contains an evolved companion, similar to \novacyg.  There are
differences in the optical spectrum: for example, \novacyg\ showed
very strong emission lines of hydrogen, helium and other species
(Casares et al.\ 1991, Wagner et al.\ 1998), whereas \xtesource\ shows
only weak H$\alpha$ emission.  This may not be significant; \novacyg\
was unusual in this respect, and most SXTs show few emission lines
during outburst.  Another long period system, \novasco, has exhibited
outbursts both with H$\alpha$ emission \cite{B95} and without
\cite{H98}.

More puzzling is the apparent plateau in the X-ray brightness after
day 95 at a level of $\sim12$\,mCrab.  In this period, the flux is
gradually rising and culminates in a mini-outburst around day 145,
approximately 120 days after the secondary maximum.  This is an
intriguing timescale as both \novasco\ (Harmon et al.\ 1995, Zhang et
al.\ 1997) and \novaper\ (Augusteijn, Kuulkers \& Shaham 1993,
Chevalier \& Ilovaisky 1995) have also shown mini-outbursts separated
by $\sim120$ days.  The lightcurves of these three systems are
otherwise very different, and it is not clear that their
mini-outbursts are caused by the same mechanism; see Chen, Shrader \&
Livio (1997) for a discussion of the models proposed for SXT
rebrightenings.  

We conclude that \xtesource\ shows similarities to other SXTs,
especially long-period systems.  The extended plateau state is,
however, unusual.  Continued monitoring is important to elucidate its
nature.  In particular it will be of interest to see if further
mini-outbursts occur.
%
%
\section*{Acknowledgements}
The Jacobus Kapteyn and William Herschel Telescopes are operated on
the island of La Palma by the Isaac Newton Group in the Spanish
Observatorio del Roque de los Muchachos of the Instituto de
Astrofisica de Canarias. The United Kingdom Infrared Telescope is
operated by the Joint Astronomy Centre on behalf of the U.K. Particle
Physics and Astronomy Research Council.  The JKT observations were
provided by Nic Walton as part of the JKT Service Programme and UKIRT
data was obtained through the UKIRT Service Programme.  \RXTE\ results
provided by the ASM/\RXTE\ teams at MIT and at the \RXTE\ SOF and GOF
at NASA's GSFC.  RIH is supported by a PPARC Research Studentship and
would like to acknowledge helpful discussion with Carole Haswell and
others at Sussex, as well as constructive criticism from our referee,
Michael Garcia.  This research has made use of the SIMBAD database,
operated at CDS, Strasbourg, France.
%
%

%
\end{document}